# Amplification of femtosecond pulses with AI-assisted spectral phase modulation

MIKOŁAJ KRAKOWSKI,[1,†] ALICJA KWAŚNY,[1,†] AND GRZEGORZ SOBOŃ[1,*]

[1]*Laser & Fiber Electronics Group, Faculty of Electronics, Photonics and Microsystems, Wroclaw University of Science and Technology, Wybrzeze Wyspianskiego 27, 50-370 Wroclaw, Poland*
[†]*These authors contributed equally*
*\*grzegorz.sobon@pwr.edu.pl*

**Abstract:** We report our investigation on ultrashort laser pulse optimization using an AI algorithm in a system consisting of a mode-locked oscillator, a spectral phase shaper, and a highly nonlinear amplifier. We analyzed the performance of the pulse optimization process as a function of two main parameters: the resolution of spectral phase modulation and the number of agents in the algorithm. We showed that the algorithm could find an optimum phase profile for the seed pulse, which allowed for a reduction of the FWHM of the amplified pulse by 10 fs (from 46 to 36 fs), and significantly reduced the intensity of the side-pulse by a factor of 4.6. Importantly, the algorithm used does not require any training and optimizes the pulse shape without any knowledge about the input pulse parameters or the parameters of the amplifier. We believe the proposed system might be a convenient test bed for evaluating various AI-based algorithms in a pulse optimization task.

## 1. Introduction

Generation of ultrashort laser pulses, with duration at the level of a few optical cycles, is critical for multiple applications, like electro-optical sampling [1], attosecond pulse generation [2], THz generation [3], ultrafast time-resolved spectroscopy [4], and others. In most of these applications, the system's front end is a mode-locked laser emitting femtosecond pulses, which are subsequently appropriately modified (amplified, compressed, or converted to a different spectral range) to meet the requirement of the application. Recently, a novel approach for designing ultrafast laser sources, referred to as "smart photonics", has emerged [5,6]. In such an approach, the optical system is equipped with electronically controlled components that allow for changing the parameters of the output pulse (e.g., a tunable filter, rotating waveplate, etc.), and machine learning algorithms are used to find the optimum settings and obtain the desired parameters.

So far, machine learning has been mostly applied for optimizing and self-tuning mode-locked fiber lasers (oscillators) [7–15]. Those based on nonlinear polarization rotation (NPR) are an ideal platform for investigating AI-based optimization methods. An NPR-based oscillator usually contains two or three wave plates ($\lambda/4$ and $\lambda/2$). The pulsed operation is achieved by finding the proper settings of the wave plates angles. The temporal and spectral parameters of the output pulse critically depend on these settings, since the properties of the artificial saturable absorber are polarization-dependent. Moreover, the optical fibers in NPR-based lasers are non-polarization maintaining (non-PM). Therefore, finding the optimum settings is non-repeatable and time-consuming. In 2013, Fu and Kutz proposed to combine a feedback loop with a genetic algorithm to control the cavity parameters of an NPR-based laser [16]. Later, many experiments were carried out in various configurations, dispersion regimes, and types of AI used for optimization. It was shown that solutions based on genetic or evolutionary algorithms allow the optimized pulse to be found within less than 30 minutes [5,7], which usually takes hours or days of lab work. All these works have proven that machine learning algorithms are helpful in optimizing lasers - the algorithm might find much better settings in a broad parameter space than humans; moreover, they can do it much faster.

Recently, a straightforward approach to generate few-cycle pulses in the spectral band of 1.56 μm, based on nonlinear amplification, became highly popular among scientists [17]. Such scheme consists of a mode-locked oscillator, followed by an Erbium-doped fiber amplifier (often in an all-fiber configuration), with no grating-based stretchers or compressors. The pulses from the oscillator might be relatively long (at the order of a few hundreds of femtoseconds). In contrast to classical chirped pulse amplification (CPA), the pulses experience excessive spectral broadening during the amplification process, which enables their compression to have very short durations. In the most straightforward possible scheme, such a system can consist only of two types of fibers (single-mode fiber and the Er-doped fiber) [18]. Proper matching of the length of these two fibers can lead to the generation of very short pulses, at the level of few optical cycles. More complex setups include more sophisticated ways to control the chirp of the pulses, by means of prism pairs [19–23], or a combination of optical fibers (including negative dispersion fibers, dispersion-shifted fibers, photonic crystal fibers, or highly-nonlinear fibers [17,24–28].

The main difficulty in optimizing a particular ultrafast laser system (consisting of an oscillator and an amplifier) arises from the number of degrees of freedom that must be balanced to achieve the desired parameters (for example, the shortest pulse duration at the output). A most straightforward method that could be applied to auto-tune a laser system would be a linear sweep through the values of a given parameter while monitoring the system output, and using a feedback loop to obtain (and maintain) a desired pulse shape. However, such an approach becomes useless in the abovementioned nonlinear amplifiers, where the final pulse shape depends on many degrees of freedom, and many parameters must be optimized simultaneously (e.g., lengths of all individual fibers, settings of the prisms, powers of the pump lasers). The shortest output pulse duration is achieved only at one particular setting ("sweet spot") of all the degrees of freedom. Moreover, in a typical arrangement of a nonlinear amplifier, the output pulse shape depends not only on the parameters of the amplifier (its dispersion, nonlinearity, gain), but also on the input pulse parameters (spectral width, temporal width, spectral phase, peak power). Thus, any minor change to the oscillator will destroy the final output pulse shape and require re-arrangement of the amplifier. Therefore, manual optimization of such systems in a typical laboratory experiment is challenging and time-consuming. So far, adaptive AI-based algorithms were used to control the propagation of ultrashort pulses through various media, like optical fibers [29,30] or glass rods [31]. Farfan et al. developed an adaptive neural network algorithm to control a pulse shaper based on a deformable mirror, showing that machine learning can significantly accelerate the process of pulse compression [32]. Adaptive pulse control was also applied in chirped pulse amplifiers to optimize the pulse shape [33–35]. However, we note that the previous demonstrations focused mostly on high-power systems, and were using genetic algorithms [33,36], or algorithms based on stimulated annealing [37,38].

To overcome the limitations of manual optimization of ultrafast oscillator-amplifier setups, we propose using arbitrary phase shaping of the pulses combined with an AI-based algorithm for self-optimizing the laser system. In our experiment, we placed a programmable phase modulator between the oscillator and the amplifier. The modulator was feedback-controlled by a metaheuristic algorithm, Grey Wolf Optimizer. The algorithm generates an optimal phase mask for the modulator based on the measured autocorrelation trace. Notably, the algorithm does not require any training prior to operation, and optimizes the pulse shape without any knowledge about the input pulse parameters or the parameters of the amplifier. We show that the algorithm can find the optimum pulse shape in several minutes, significantly improving the pulse duration and quality by minimizing the side-pulses. The proposed system is an excellent test bed for evaluating various AI-based algorithms in a pulse optimization task.

## 2. Experiment

### 2.1 Laser and amplifier setup

Figure 1 shows the experimental setup of the AI-optimized laser system. As a seed source, a ring cavity Er-doped fiber laser (EDFL) mode-locked via semiconductor saturable absorber mirror (SESAM) was used (Batop GmbH). The seed generated 560-fs pulses at a 50 MHz repetition rate and an average power of 4.3 mW at 1560 nm central wavelength. The pulses from the seed were dispersed on a diffraction grating with 1100 lines/mm (Shimadzu Corp.) and directed to a spatial light modulator (Santec SLM-200, SLM). The free-space part was arranged in a 4-f configuration which induced no additional dispersion. Next, the pulses were coupled back to an optical fiber and pre-amplified in an in-house made Erbium-doped fiber amplifier (EDFA) based on a 1.27 m-long segment of highly-Erbium doped fiber (nLight Liekki Er80-4/125-PM, EDF) pumped with a 976 nm single-mode laser diode with output power of 1W (II-VI Coherent CM97-1000-76PM). The amplified pulses were recorded with an autocorrelator (APE PulseCheck NX50, AC) connected to a PC (Acer Nitro 5). An in-house made Python-based software was used to acquire the data from the AC and compute a phase mask sent to the SLM. The SLM acts as a display to a PC, in which the phase of each individual pixel can be changed from $0\pi$ to $2\pi$ in a 10-bit depth range (1024 gray levels).

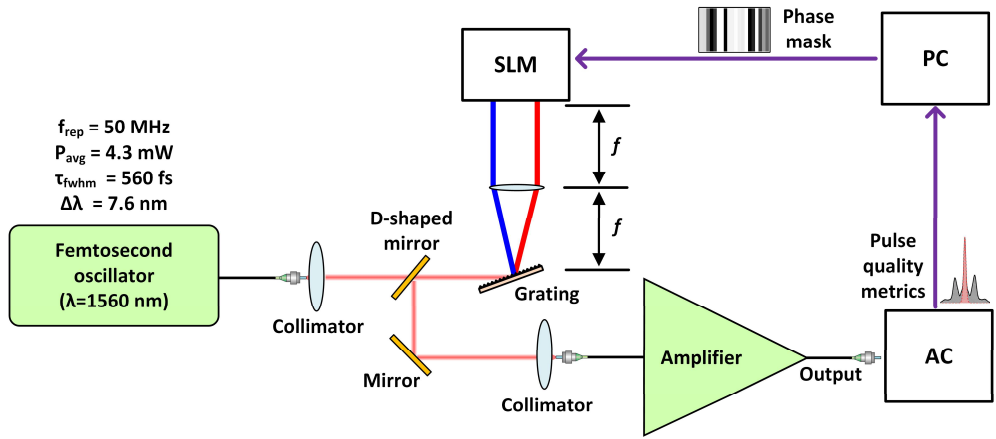

Fig. 1. Experimental setup. AC – Autocorrelator, PC – Personal Computer, SLM – Spatial Light Modulator.

The setup was entirely fiberized and spliced using a standard arc-fusion splicer. All fibers and components used in this setup were single-mode and polarization-maintaining (PM-SMF). The entire system was built using only two types of fibers (the EDF and PM-SMF). All components are standard, off-the-shelf PM fiber components (Advanced Fiber Resources).

### 2.2 Grey-Wolf Optimizer

The algorithm that was used to optimize the pulses was the Grey Wolf Optimizer [39]. It has already shown its applicability in the optimization of electromagnetic devices [40], power flow problems, or in photonic crystal filter design [41]. It belongs to the category of meta-heuristic algorithms, which search for optimal solutions by iteratively exploring and exploiting the solution space. Many of them are inspired by natural phenomena: the Grey Wolf Optimizer, mimics the hunting behavior of a pack of wolves. Notably, the GWO is an online machine learning algorithm that learns in real time and does not need any prior data acquisition. As it does not use any neurons, or deep learning architectures, it does not require powerful computational resources (a regular PC is sufficient).

The pseudo-code shown in Fig. 2 shows a simplified implementation of the GWO algorithm. In order to find the optimal phase mask (corresponding to the best laser pulse), we create a pack of wolves (agents), and the three fittest solutions (α, β and δ). Following the behavior of the pack, mechanisms of encircling prey and hunting occur in the algorithm. The encircling is modeled using equations (1) and (2):

$$\vec{D} = |\vec{C}.\vec{X_p}(t) - \vec{X}(t)|, \tag{1}$$

$$\vec{X}(t+1) = \vec{X_p}(t) - \vec{A}.\vec{D}, \tag{2}$$

where $t$ is the iteration, $\vec{A}$ and $\vec{C}$ are coefficient vectors, $\vec{X_p}$ is the position vector of the prey (global minimum) and $\vec{X}$ is the position of the grey wolf (agent). Vectors $\vec{A}$ and $\vec{C}$ introduce randomness to the algorithm, and are determined in the following manner:

$$\vec{A} = 2\vec{a}.\vec{r_1} - \vec{a}, \tag{3}$$

$$\vec{C} = 2.\vec{r_2}, \tag{4}$$

where components of $\vec{a}$ undergo a linear decrease from 2 to 0 throughout the iterations and $r_1$, $r_2$ represent random vectors within the [0,1] range. In each iteration of the algorithm, we generate a chosen number of agents and check each of the generated positions by displaying the corresponding SLM mask, acquiring the autocorrelation function, and calculating its fitness. We then compare the fitness with our three best solutions (α, β, and δ) and if it is better, we replace the value of α, β or δ with the new position. Parameter A is also introduced, as a exploration-exploitation mechanism. It changes from 1 to 0 throughout the iterations. Finally, the solution is the mask corresponding to the α solution. The parameters that can be adjusted influence the behavior of the algorithm over iterations are: the number of agents (results in more thorough search in each iteration), the number of stripes (increases the spectral resolution but might require more iterations to achieve the global minimum), and the number of iterations.

```
Initialize the grey wolf population X_i (i = 1, 2, ..., n)
Initialize a, A, and C
for each search agent
    Display on SLM the mask corresponding to search agent
    Read the autocorrelation function
    Calculate the fitness of the search agent
end for
X_α = the best search agent
X_β = the second best search agent
X_δ = the third best search agent
while (t<Max number of iterations)
    for each search agent
        Display on SLM the mask corresponding to search agent
        Read the autocorrelation function
        Calculate the fitness of the search agent
        Update X_α, X_β and X_δ
    end for
    Update a, A, and C
    t = t+1
end while
return X_α
```

Fig. 2. Pseudo code of the implementation of the GWO algorithm (based on Ref. [25]).

Each optimization algorithm needs a well-defined optimization goal (cost function). In the case of optimizing laser pulses, the most straightforward cost function could be simply the pulse duration measured by an autocorrelator. However, in the case of highly nonlinear amplification, the pulses often contain a pedestal or pre-/post-pulses in the AC trace. In other words, the main pulse might be very short, but the energy might be stored in the pedestal or side pulses. To avoid such situation we proposed a new pulse quality factor that takes into account the FWHM, but also the shape of the pulse (closeness to the theoretical sech2 fit), defined as:

$$F = FWHM \cdot (1 + \left|\frac{A_{fit}}{A}\right|)^2, \tag{5}$$

where FWHM is the full-width half maximum of the AC function of the pulse, $A_{fit}$ is the area under the sech$^2$ fit of the pulse, and A is the area of the AC trace. The part of the equation in brackets is squared, in order to put more emphasis on the shape of the pulse.

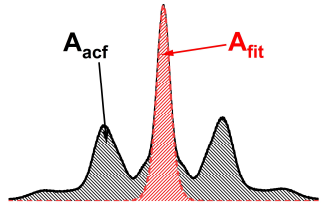

Fig. 3. Representation of areas under sech$^2$ fit (red) and autocorrelation function (black).

## 3. Experimental results

The performance of the laser system in all experiments was observed using an optical spectrum analyzer (Yokogawa AQ6375, OSA), an optical autocorrelator (APE PulseCheck NX50, AC), and a frequency-resolved optical gating device (Mesa Photonics FS-Ultra2, FROG).

### 3.1. Non-optimized system

Firstly, we characterized the system in a neutral state, i.e., without any phase mask applied to the SLM, as a reference for the experiments with optimization. Figure 4(a) shows the optical spectrum with the spectral phase. It can be seen that the spectrum experiences significant spectral broadening. The temporal profile of the pulse, together with the temporal phase is depicted in Fig. 4(b). The FHWM pulse duration was 46 fs, and the profile contained significant side-pulses. The intensity of the first side pulse in the FROG trace equals 0.37 of the intensity of the main peak. We estimated from the FROG trace that the side-pulses contain approximately 28% of the total energy of the pulse. Fig. 4(c) shows the recorded AC trace with the sech$^2$ fit.

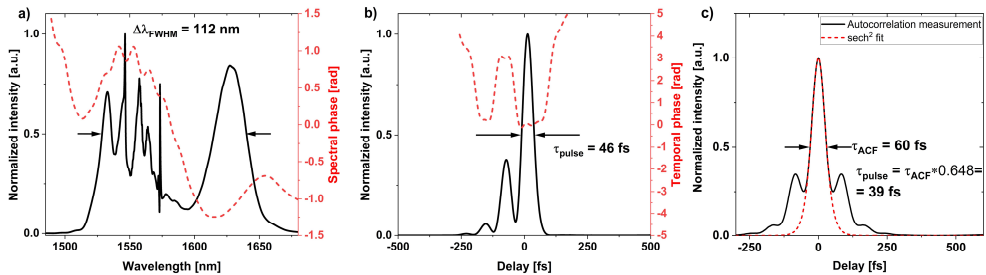

Fig. 4. Characterization of the non-optimized (neutral) system: a) optical spectrum (solid black line) with the spectral phase (dashed red line); b) temporal intensity profile of the pulse (solid black line) with the temporal phase (dashed red line); c) autocorrelation trace of the pulse (solid black line) with the sech$^2$ fit (dashed red line).

## 3.2. Pulse optimization using GWO

In the experiment, we investigated the performance of the optimization algorithm as a function of two main parameters: the spectral resolution (i.e., the width and number of stripes displayed on the SLM), and the number of agents. The number of iterations was fixed in all measurements and set to 20. We used the following number of stripes (n) displayed on the SLM: 20, 40, 80, 160, 320 and 480. For each setting, we ran the algorithm with 50, 150, 300, 500, and 800 agents. We recorded the obtained pulse shape and the time required for the algorithm to accomplish 20 iterations. The pump power of the amplifier remained the same in all experiments and equaled 1W. Figure 5 illustrates the way the phase mask was displayed on the SLM. The operation has been simplified by assuming the beam to be 1-dimensional; thus the division of the X (width) axis into stripes of defined length (in pixels and μm) reduces complexity and optimization time.

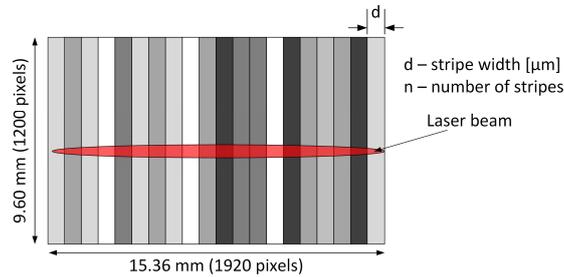

Fig. 5. Simplified depiction of the mask display and beam placement on the SLM. In the presented case the number of stripes: n=20, and the corresponding stripe width d=768 μm.

Table 1 summarizes the parameters of the phase masks displayed on the SLM. The highest investigated spectral resolution was 0.075 nm, corresponding to a stripe size of 4 pixels (corresponding to a physical width of 32 μm). In this case, the SLM is divided to 480 stripes. The lowest resolution in our experiments was 1.8 nm (stripe width d=768 μm, number of stripes n=20).

Table 1. Definition of parameters used in optimization.

| Number of stripes (n) | Stripe width in pixels | Stripe width (d) [μm] | Spectral resolution [nm] |
| --- | --- | --- | --- |
| 20 | 96 | 768 | 1.8 |
| 40 | 48 | 384 | 0.9 |
| 80 | 24 | 192 | 0.45 |
| 160 | 12 | 96 | 0.225 |
| 320 | 6 | 48 | 0.1125 |
| 480 | 4 | 32 | 0.075 |

In order to better compare and understand the behavior of our algorithm with different parameter values, we drew fitness characteristics for all investigated cases, shown in Fig. 6. Such characteristic shows how the algorithm behaved throughout the iterations, e.g., whether it got stuck in a local minimum. It can be seen that increasing the number of stripes increases the computational complexity, and the algorithm requires more iterations to achieve the minimum cost function. For example, the cost function did not reach the minimum with 480 stripes and more than 300 agents. The best output pulse was achieved for 80 stripes and 800 agents (Fig. 5d): the fitness score was equal to ~139 fs, which corresponded to a pulse duration of ~36 fs. Interestingly, increasing the spectral resolution (i.e., increasing the number of stripes) does not improve the final pulse quality. Our study shows that the resolution of ~0.45 nm (stripe width of 192 μm) gives the best result, but even 1.8 nm resolution is sufficient to improve the pulse shape significantly.

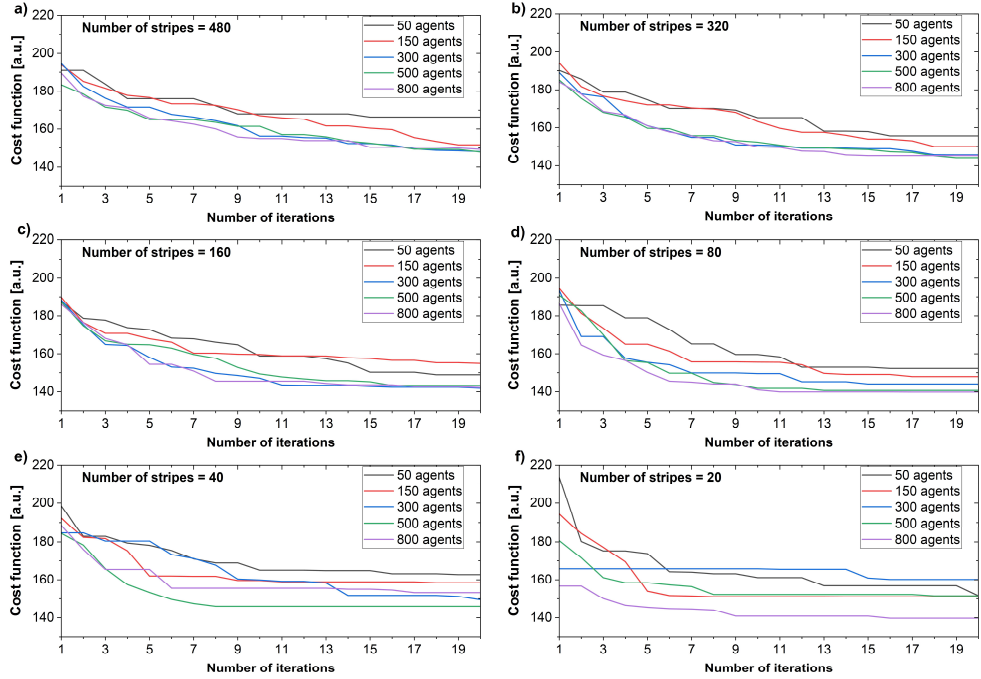

Fig. 6. Cost functions for constant number of iterations – 20; the varying parameters are the number of stripes on the SLM and number of the agents in the GWO algorithm.

One of the most critical parameters of the optimization is the time of the operation. In GWO, the optimization time depends on the number of agents and iterations. We measured the time required to accomplish 20 iterations in all investigated cases. The results are summarized in Table 2. We want to clearly emphasize that the limitation in the computation time was not the GWO algorithm itself or the computing power of the PC. The main limiting factor here is the data acquisition from the autocorrelator. Due to the communication interface with the device, we were able to collect AC traces every second, while the calculation of one phase mask takes approximately only 10 μs. Replacing the used autocorrelator with a different device with much faster acquisition rates would dramatically improve the time required for the calculations.

Table 2. Time required for accomplishing 20 iterations.

| No. of agents: | 50 | 150 | 300 | 500 | 800 |
|---|---|---|---|---|---|
| Optimization time [min]: | 22 | 66 | 132 | 220 | 353 |

But what puts it in the wider perspective is the correlation between the number of stripes, the number of agents, and the number of iterations required to reach the minimum fitness score. Comparing Fig. 5b (320 stripes) and Fig. 5e (40 stripes) we observe that both the optimization reached similar fitness score, but for the smaller number of stripes the curve flattens much quicker. We can dispute if the 320 stripe case would reach lower fitness score value if handed additional iterations (longer time of optimization). However, we decided to limit the number of iterations to 20 to arrive at the optimum point at a reasonable time. From this, we can conclude that the higher the resolution, the longer the search for the optimal point.

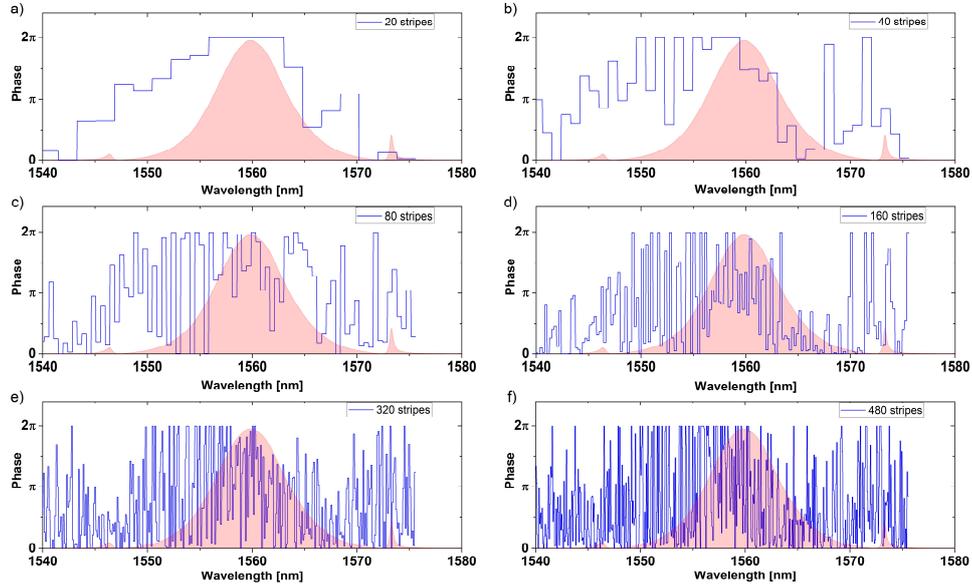

Fig. 7. Generated phase masks in GWO settings of 800 agents, for a) 20, b) 40, c) 80, d) 160, e) 320, and f) 480 stripes. The red area in the background illustrates the shape of the oscillator spectrum.

Finally, we have performed a detailed characterization of the most optimal pulse, which was achieved with 80 stripes and 800 agents (mask shown in Fig. 7c), which proved to reach the fitness score of 139.9 fs. Figure 8a shows the optical spectrum with the spectral phase. The temporal profile of the pulse, together with the temporal phase is depicted in Fig. 8(b), and Fig. 8(c) shows the recorded AC trace with the sech$^2$ fit.

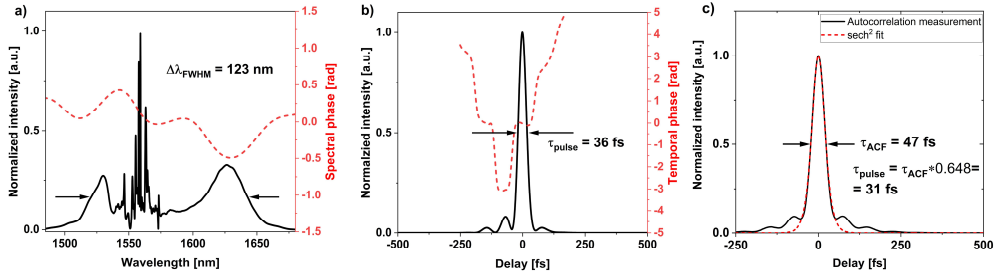

Fig. 8. Characterization of the GWO-optimized pulse: a) optical spectrum (solid black line) with the spectral phase (dashed red line); b) temporal intensity profile of the pulse (solid black line) with the temporal phase (dashed red line); c) autocorrelation of the pulse (solid black line) with the sech$^2$ fit (dashed red line).

Comparing the measurements of the optimized to the non-optimized pulse we observed a significant suppression of the side-pulse intensity and shortening of the pulse FWHM by 10 fs (from 46 to 36 fs). The intensity of the first side-pulse related to the intensity of the main peak was only 0.08 (reduction by a factor of 4.6). We estimate that the side-pulses contain approximately only 10% of the total energy of the pulse. The spectrum of the pulse has widened by 11 nm (from 112 to 123 nm). The spectral phase of the optimized pulse has a remarkably smaller variation (nearly two times smaller).

## 4. Summary

In summary, we have presented our investigation of arbitrary phase shaping of laser pulses prior to amplification, combined with an AI-based algorithm for self-optimizing the laser system. In our experiments, we evaluated the performance of the GWO algorithm, which is one of the emerging AI algorithms used to solve practical engineering problems. We analyzed the performance of the pulse optimization process as a function of two main parameters: the resolution of spectral phase modulation, and the number of agents in the GWO algorithm. We showed that the algorithm is capable of finding an optimum phase profile for the pulse, which allowed to reduce the FWHM of the pulse by 10 fs (from 46 to 36 fs), and significantly reduce the intensity of the side-pulse by a factor of 4.6. We also conclude that the spectral resolution of the phase modulation does not play a crucial role in the optimization: the best performance was achieved with a resolution of 0.45 nm, and further increase of the resolution did not provide better results. In the future, we plan to use the presented system as a test bed for exploring other AI-based methods, such as genetic algorithm or crow search.

We emphasize that the optimization speed was strongly limited by the communication with the autocorrelator. In order to improve the optimization speed we plan to investigate other types of feedback signals, e.g., directly measuring a two-photon response of a photodetector, or using an autocorrelator with much faster acquisition rates.

**Funding.** Narodowe Centrum Nauki (2021/42/E/ST7/00111).

**Acknowledgments.** This research was funded in whole by the National Science Centre (NCN) under the grant no. 2021/42/E/ST7/00111. For the purpose of Open Access, the author has applied a CC-BY public copyright license to any Author Accepted Manuscript (AAM) version arising from this submission.

**Disclosures.** The authors declare no competing interests.

**Data availability.** Data underlying the results presented in this paper are available in the Dataset under Ref. [42].